\documentclass[DIV=12]{article}
\usepackage[margin=1.1in]{geometry}
\usepackage{subfig}

\usepackage[dvips]{color}
\usepackage{multido}
\usepackage{graphicx}
\usepackage{amsmath}
\usepackage{tabularx}

\newcommand{\bigTwelve}{{\ttfamily{'big12'}}}

\begin{document}

\title{
\noindent\hrulefill
\begin{flushleft}
{\Large \bf{Cell-type-specific computational neuroanatomy, simulations from the sagittal and coronal Allen Brain Atlas\\
\vspace{8mm}
\hrule}}
\vspace{8mm} 
{\large{Pascal Grange\\
Xi'an Jiaotong--Liverpool University, Department of Mathematical Sciences\\
 111 Ren'ai Rd, Science Building SD557, Suzhou 215123, Jiangsu Province, China\\
{\ttfamily{pascal.grange@xjtlu.edu.cn}}\\
}}
\noindent\hrulefill
\end{flushleft}}
\date{}
\author{}
\maketitle
\vspace{-5mm}
 
\begin{abstract}

The Allen Atlas of the adult mouse brain is a brain-wide, genome-wide data set
  that has been made available online, triggering a renaissance in neuroanatomy. In particular,
  it has been used to define brain regions in a computational, data-driven way, and to estimate the region-specificity 
 of cell types characterized independently by their transcriptional activity. However, these results 
  were based on one series of co-registered (coronal) ISH image series per gene, whereas the online 
 ABA contains several image series per genes, including sagittal ones. Since the sagittal series
 cover mostly the left hemisphere, we can simulate the variability of results by repeatedly drawing a random 
 image series for each gene and restricting the computation to the left hemisphere. This gives
 rise to an estimate of error bars on the results of computational neuroanatomy. 
 This manuscript is a complement to the paper {\emph{Computational neuroanatomy: mapping cell-type densities in the mouse brain, 
simulations from the Allen Brain Atlas}} prepared for the International Conference on Mathematical Modeling in Physical Sciences June 5-8, 2015,
 Mykonos Island, Greece.

\end{abstract}

\clearpage


{\bf{Acronyms.}} ABA: Allen Brain Atlas; ARA: Allen Reference Atlas; ISH: {\emph{in situ}} hybridization.\\
\vspace{3mm}

\tableofcontents 

\section{Introduction}

%
%

 The Allen Brain Atlas (ABA, \cite{AllenGenome,AllenAtlasMol,corrStructureAllen}) has renewed the gene-based
 approach to gene-expression studies in neuroscience by releasing voxelized, 
 brain-wide ISH data for the entire genome of the mouse, which were co-registered to 
the Allen Reference Atlas (ARA, \cite{ARA}).  About 4,000 genes of special neurobiological interest 
  were proritized. For these genes 
 an entire brain was sliced coronally and processed (giving rise to the coronal ABA). For the rest of the genome 
 the brain was sliced sagitally, and only the left hemisphere was processed (giving rise to the sagittal ABA). In the computational
 approaches of \cite{corrStructureAllen,markers,qbCoExpr,autismCoExpr,cellTypeBased,autismTypes}, only data from the coronal ABA were
   analyzed, in order to obtain brain-wide results. However, these brain-wide results revealed a large degree
 of left-right symmetry. Moreover, more than 90\% of the genome was observed in \cite{AllenGenome,AllenAtlasMol}
 to be expressed in the brain. There is therefore a strong need to incorporate data from the sagittal atlas
 into the analysis of the ABA. The website {\ttfamily{www.mouse-brain.org}} already caters to this need for each gene,
 because a query based on a gene name returns all the ISH image series for the corresponding gene (specifying whether 
 it comes from sagittal or coronal sections), together with a bar diagram summarizing how the gene expression
 breaks up between the main regions of the brain. This enables a user of the atlas to compare the expression profiles
 coming from all the available data, on a gene-by-gene basis. The results are reproducible on a desktop computer
 using the Brain Gene Expression Analysis MATLAB toolbox \cite{BGEA}.\\

  The ABA has  proven to be a powerful tool for systems biology, because co-registered 
 gene-expression data can be studied collectively, thousands of genes at a time.
 Indeed the collective behaviour of gene-expression data is crucial for the analysis of \cite{cellTypeBased,supplementary1,supplementary2},
 in which the brain-wide correlation between the ABA and cell-type-specific microarray data was studied.
 Strong heterogeneities of the correlation profiles were observed, allowing to guess,
 for instance that medium spiny neurons were extracted from the caudoputamen. Moreover, 
 the region-specificity of cell types was estimated by linear regression with 
  positivity constraint. The model was fitted using the coronal atlas only, which allowed to 
 obtain brain-wide results. However, this restriction implies the availability
 of only one ISH gene-expression profile per gene. This poses the problem 
 of the error bars on the results of the model.\\

 In the present paper we shall extend the analysis to include ISH data from the sagittal atlas.
 These data are based on the sections of the left hemisphere only, which induces
 a restriction of neuroanatomical results to the left hemisphere, but allows to simulate 
 the distribution of analysis results involving the collective behaviour
 of gene-expression profiles {\emph{without restricting the number of genes}}.
 Since the ISH data in the sagittal atlas were co-registered to the 
 voxelized ARA (at a resolution of 200 microns), they can be treated computationally in the same way as the coronal
 atlas.

\section{Methods}
\subsection{Gene expression data from the ABA}
 The digitized expression energies obtained by measuring the grey-scale intensities 
 of ISH image series co-registered to the ARA (\cite{AllenGenome,AllenAtlasMol})
 can be arranged into matrices, in which rows correspond to voxels
 and columns to genes. At a spatial resolution of 200 microns, there are $V=49,742$
 cubix voxels in the mouse brain).
 Consider a gene labeled $g$ in the coronal atlas, and a voxel
 labeled $v$ in the mouse brain. This gene
 comes with a number of gene-expression profiles 
 in the ABA, call this number $N(g)$. For most genes in the 
 coronal atlas $N(g) = 2$, with one coronal profile and one sagittal profile,
 but there are cases with more data. Let us label these ISH profiles
 by integers between $1$ and $N(g)$ (with the label $1$ corresponding 
 to the coronal atlas, for definiteness). Let us denote by $E^{(k)}(v,g)$ the gene-expression energy of 
 gene $g$ at voxel $v$, so that $E^{(1)}(v,g)$ is positive number corresponding to 
 the gene-expression energy of gene $g$ at voxel $v$ in the coronal atlas\footnote{This is the quantity denoted by $E(v,g)$ in
 papers \cite{methodsPaper,markers,BGEA,autismCoExpr,cellTypeBased}}.
  For higher values of the index $k$, the entry $E^{(k)}(v,g)$ is a positive
 number if voxel labeled $v$ is covered in the tissue sections
 of sample labeled $k$, and is labeled as missing data otherwise (which is 
 the case if the label $k$ corresponds to a sagittal sample,
 and voxel $v$ is in the right hemisphere).\\

\begin{figure}
\centering
      \includegraphics[width=0.9\textwidth]{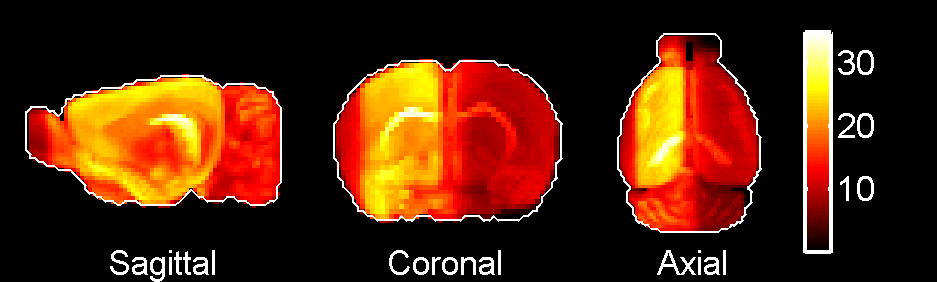}
    \caption{Maximal-intensity projections of the sum of all gene-expression profiles in the atlas (from both sagittal and coronal series).
 The left hemisphere has higher intensity then the right hemisphere due to the larger number of series covering it. However, the anatomical heterogeneity 
 patterns (highlighting the hippocampus for instance), apart from intensity, seem to be left-right symmetric. However, there is a darker area on the 
left hemisphere that reveals missing sections from the sagittal series: indeed in this region the slices ae extremely small and
 fragile. The same issue is encountered in the most frontal sections of the olfactory buln in the coronal series.}
    \label{intersectionDef}
  \end{figure}

 When working with all the image series in the atlas at the same time, we have 
 to restrict to voxels that are in the intersection of the 
 sets of voxels for which all genes have been processed
 A maximal-intensity projection of the sum of all gene-expression profile shown on Fig. \ref{intersectionDef},
 which shows that the data are not homogeneous
 over the brain hemisphere on average, and that the left hemisphere indeed has a larger signal 
 than the right hemisphere, because of the larger number of image series obtained
 in the ABA (the voxels with no data were treated as zero entries when performing the sum of gene expression energies).\\

\subsection{Simulation of variability}

 For computational purposes 
 we have the choice between $N(g)$  expression profiles for gene labeled $g$. 
 Hence, instead of just one voxel-by-gene matrix whose $g$-th column 
 contains $\left(E^{(1)}(v,g)\right)_{1\leq v \leq V}$,
 the ABA gives rise to a family of $\prod_{g=1}^G N(g)$ voxel-by-gene matrices, with voxels belonging to the 
  left hemisphere. Any quantity computed from 
 $E^{(1)}$  can be recomputed from any of these matrices, thereby inducing
 a distribution for this quantity, from which we can 
  estimate quantities such as mean, median, variance and density. This is a finite but prohibitively large number of
 computations (at least $2^G$, given that each gene in the coronal atlas has at least two image series) to be performed, so
 for practical purposes 
 we have to take a Monte Carlo 
 approach and to perform the computation for $R$ random choices of data. The simulation of the 
 distribution of a quantity $Q$ depending on the atlas can be obtained by the following pseudo-code:\\
{\ttfamily{
for all $i$ in $[1..R]$}}\\
{\ttfamily{1. for all $g$ in [1..G], choose an integer $n_i(g)$ in $[1.. N(g)]$}}\\
{\ttfamily{2. construct the matrix $E_{[i]}$ with entries $E_{[i]}(v,g) =E^{(n_i(g))}(v,g)$}}\\
{\ttfamily{3. compute the quantity $Q$ using this matrix, call the result $Q_{[i]}$}}\\
{\ttfamily{end\\
}}
 The larger $R$ is, the more precise the estimates for the distribution of the quantity $Q$ will be.\\

\subsection{Correlation between the ABA and cell-type-specific microarray data}

As an application, consider the correlation profile between the ABA and cell-type-specific
 transcriptome profiles collated in \cite{OkatyComparison}. The microarray reads 
 are arranged into a matrix $C$, whose rows correspond to cell types and whose columns correspond to genes (arranged
 in the same orders as in the matrix presentation of the ABA expression energies): $C(t,g)$ is the microarray read for
 gene labeled $g$ in the cell-type-specific sample labeled $t$, in the notations of \cite{cellTypeBased},
 were the index $t$ take values between $1$ and $T=64$, and $g$ takes values getween $1$ and $G=2,131$,
 which is the number of genes found in \cite{cellTypeBased} to be found both in the coronal ABA and in all the microarray
 reads).
 If we represent the ABA  by a matrix $E_{[i]}$ of ISH data in the left hemisphere, obtained by drawing 
 a random set in step 2 of the pseudo-code, the value of the brain-wide correlation at voxel $v$ with cell type $t$ reads:
\begin{equation}
\mathrm{Corr}_{[i]}(v,t)=\frac{\sum_{g=1}^G\left(C(t,g)-\frac{1}{T}\sum_{t=1}^T C(t,g)\right)\left( E_{[i]}( v,g )-\frac{1}{V}\sum_{v=1}^VE_{[i]}(v,g)\right)}
{\sqrt{\sum_{k=1}^G\left(C(t,k)-\frac{1}{T}\sum_{t=1}^T C(t,k)\right)^2 \sum_{h=1}^G\left(E_{[i]}( v,h )-\frac{1}{V}\sum_{v=1}^V E_{[i]}(v,h) \right)^2}},
\label{corrEquation}
\end{equation}
 and $v$ takes values corresponding to the left hemisphere. 
Having computed the family of correlation matrices $(\mathrm{Corr}_{[i]})_{1\leq i \leq R}$, we can 
  study the distribution of their $t$-th column for each $t$. For example we can compute the average
 correlation profile (where the average is performed over the index $i$ labling the choice of image series):
\begin{equation}
\langle   \mathrm{Corr} (v,t) \rangle =  \frac{1}{R}\sum_{i=1}^R \mathrm{Corr}_{[i]}(v,t).
\label{meanCorrelDef}
\end{equation}

 Moreover, we can estimate the dispersion of the correlation 
 values for a cell type labeled $t$ and a region labeled $V_r$ in the ARA.
 For each random  family of data (labeled by index $i$), we can compute
 the average correlation in the voxels belonging to region $V_r$ according 
 to the ARA (this time the average is performed over voxels and the index of image series is not summed over):
\begin{equation}
 \mathrm{Corr}_{r,[i]}(t)  =  \frac{1}{|V_r|}\sum_{v\in V_r} \mathrm{Corr}_{[i]}(v,t).
\label{meanCorrelRegDef}
\end{equation}
 For a fixed value of $t$, this computation furnishes us with one family of $R$ numbers in the interval $[-1,1]$ 
 for each region in the ARA. For the correlation analyis of \cite{cellTypeBased} to be stable under a change of animal
 and sectioning modality,
 these families of numbers should come from a probability density that presents a peak. 
  \subsection{Density estimates of cell types}
The linear model proposed in \cite{cellTypeBased} estimates the density profiles
 of cell types characterized by the transcriptome profile, assuming the expression energy of each gene
 is proportional to the number of mRNA molecules at each voxel, and that the microarray read 
 of each gene in a cell type is also proportional to the number of mRNAs for this gene in this cell type. The 
 expression energy at voxel labeled $v$ must be a a sum of cell-type-specific microarray reads,
 weighted by the density of each type at each voxel:
\begin{equation}
E(v,g) = \sum_{t}\rho_t(v) C(t,g),
\label{linearModel}
\end{equation}
 where index $t$ labels cell types. In \cite{cellTypeBased}, we estimated 
 the density profiles $(\rho_t(v))$ for all voxels $v$ in the mouse brain and
 for the cell types belonging to the data set collated in \cite{OkatyComparison}.
 This estimation process was based on minimizing the difference between 
 the l.h.s. and r.h.s. of Eq. \ref{linearModel} over all the possible positive 
 coefficients $\rho$. This optimization procedure is deterministic, 
 and relating the result to the mean density (decomposing the density
 into the sum of its mean and Gaussian noise) is 
 a difficult problem in statistics (see \cite{Meinshausen2013}).\\

Some error estimates on the value of $\rho_t(v)$ were obained in \cite{cellTypeBased,supplementary2}
  using sub-sampling techniques, 
 which involved mutilating the data by keeping only a random 10\% of the 
 coronal ABA, refitting the model, and repeating the operation. This induces
 a ranking of the cell types based on the stability of the results against sub-sampling.
 However, the set of genes on which the computation is based changes 
 for each computation, and the fraction 10\% is arbitrary (even though it is close to the 
 fraction of the genome represented by our coronal data set).
 Having integrated the sagittal data into the atlas, we can now refit the 
 model with the same sets of genes, only changing the set of image series 
 from which they come, and we can do so in $\prod_{g=1}^G N(g)$ different ways.
 The only price we have to pay for this is the restriction of the 
 results to the left hemisphere. This operation is just another
 application of the procedure outlined in the pseudo-code, with the 
 role of the quantity $Q$ played by the family of numbers $\left(\rho_t(v)\right)_{1\leq t \leq T,}$, 
 for all voxels $v$ in the left hemisphere.\\
   With the notations introduced above, we denote by $(\rho_{t,{[i]}}(v))$ the 
 density estimate obtained from the random matrix of ISH data $E_{[i]}$:
 \begin{equation}
E_{[i]}(v,g) = \sum_{t}\rho_{t,[i]}(v) C(t,g) + {\mathrm{Residual}}(v,g).
\label{linearModelRandom}
 \end{equation}
 We can group the voxels by region according to the ARA as we did for
correlations in order to compare the results to classical neuroanatomy. The average 
 density across random draws of image series for cell type lebeled $t$ reads:
\begin{equation}
\langle   \rho_t(v) \rangle = \frac{1}{R}\sum_{i=1}^R\rho_{t,[i]}(v).
\label{meanFittingDef}
\end{equation}
Since the number of cells of a given type in an extensive quantity, we compute the
 fraction of the total density contributed by each region, rather than the average density $\phi{r,[i]}(t)$ 
 in region labeled $r$ for sample labeled $i$ and cell-type labeled $t$ :
\begin{equation}
 \phi_{r,[i]}(t)  =  \frac{1}{\sum_{v\in\mathrm{left\;hemisphere}}\rho_{[i],t}(v) }\sum_{v\in V_r} \rho_{t,[i]}(v).
\label{fittingDistrDef}
\end{equation}

\section{Results}
\subsection{Distributions of correlations give rise to peaks}
\begin{figure}
   \subfloat[\label{subfig-2:dummy}]{
      \includegraphics[width=0.9\textwidth]{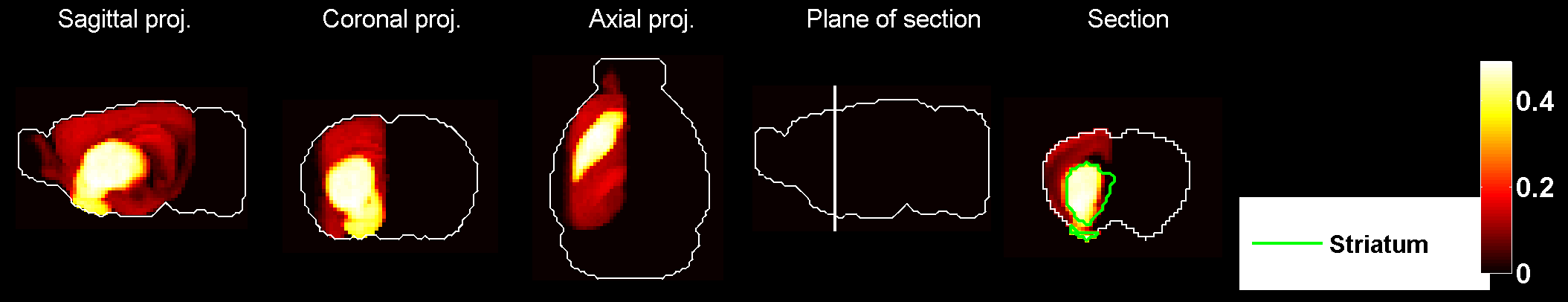}
    }
   \hfill
    \subfloat[\label{subfig-2:dummy}]{
      \includegraphics[width=0.9\textwidth]{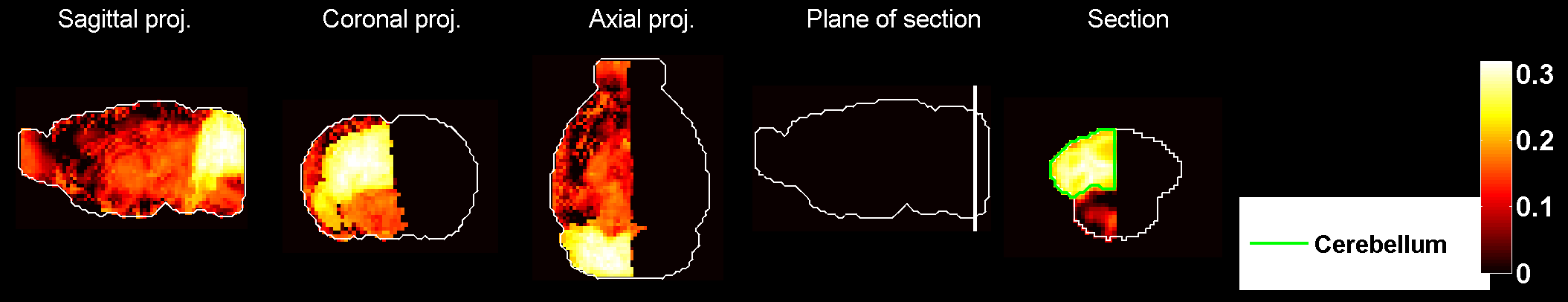}
    }
    \caption{{\bf{Heat maps of the average correlation between random image series from the ABA and cell-type-specific microarray data}},
 $\langle   \mathrm{Corr} (v,t) \rangle$, defined in Eq. \ref{meanCorrelDef}.
 (a) For medium spiny neurons, $t=16$. (b) For granule cells, $t=20$. The sections were taken through the region in the ARA in which the correlation is 
highest on average (the boundary of this region defined by the ARA is outlined in green on the section).}
    \label{meanCorrels}
  \end{figure}

\begin{figure}
   \subfloat[\label{subfig-2:dummy}]{
      \includegraphics[width=0.9\textwidth]{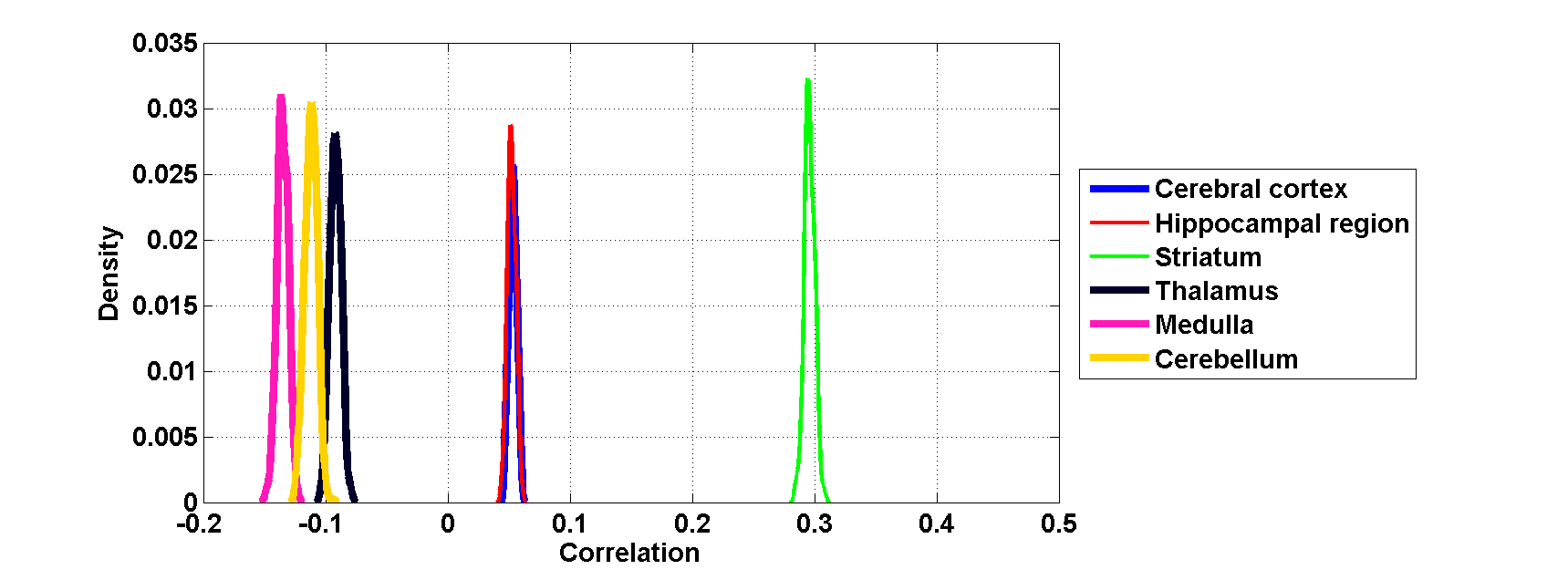}
    }
   \hfill
    \subfloat[\label{subfig-2:dummy}]{
      \includegraphics[width=0.9\textwidth]{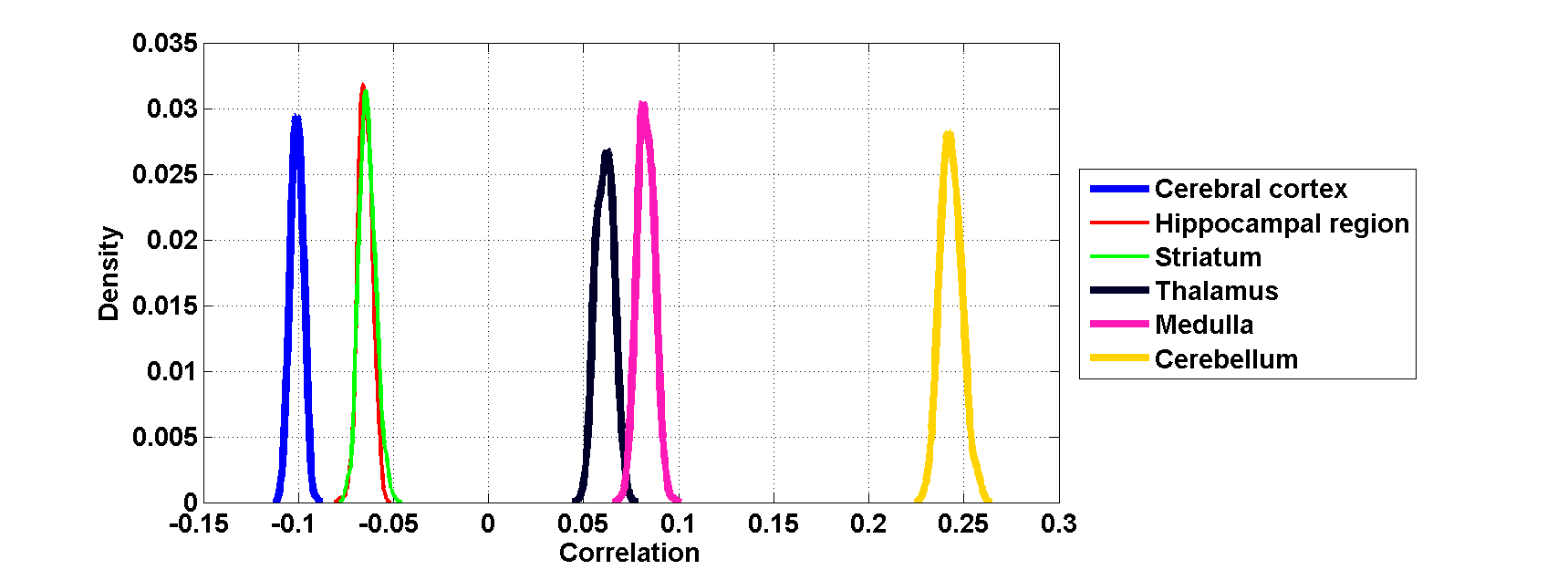}
    }
    \caption{ Estimated probability densities of the correlation profiles 
 agglomerated in a few regions of the coarsest versions of the ARA. All the region-specific densities
 give rise to peaks. The right-most peak is well-decoupled from the others in both cases.
 (a) For medium spiny neurons, $t=16$; the peak centered at the highest value corresponds to the 
 striatum. (b) For granule cells, $t=20$; the peak centered at the highest value corresponds to the 
 cerebellum.}
    \label{correlDistrs}
  \end{figure}

After running simulations,
 we computed the average correlati8on profiles defined in Eq. \ref{meanCorrelDef}.
 This quantity is plotted is plotted as a heat map on Fig. \ref{meanCorrels} for medium spiny neurons ($t=16$) and granule cells 
($t=20$), from which it is clear that the average correlation profile presents 
 a plateau inside the striatum for medium spiny neurons and inside the cerebellum for
 granule cells (which was also the case in \cite{cellTypeBased} were the correlation profiles between
 the coronal atlas and cell types were analyzed).\\
 Choosing brain regions from the coarsest version of the ARA, we computed the 
 regional averages of correlation profiles defined in Eq. \ref{meanCorrelRegDef}.
 The resulting families of numbers in the interval $[-1,1]$ were subjected
 to density estimation  (based on kernel methods in MATLAB). The estimated densities present peaks. 
Moreover, the peaks 
 are localized at higher values of correlation 
 for the regions that present striking correlation patterns. Results of density estimation 
  for medium spiny neurons ($t=16$) and granule cells 
($t=20$) are presented on Fig. \ref{correlDistrs}. In both cases the densities are peak-shaped, and the peak corresponding to the highest
 average correlation corresponds to the expected region, and is well decoupled from the 
 other peaks: the peak corresponding to the striatum for the medium spiny neurons is centered at $0.29$, and entirely supported
 in the interval $[0.27, 0.31]$, in which none of the quantities $ \mathrm{Corr}_{r,[i]}(16)$ is found for values of $r$ not corresponding to the striatum.
 These results induce reassuring bounds pointing at the stability of the 
 results of \cite{cellTypeBased,supplementary1}, even though the dark areas in the coronal and axial projections of Fig. \ref{meanCorrels}a
 probably corresponds to an artifically low expression energy in the most 
 lateral part of the left hemisphere, due to missing sections in sagittal series. This effect is probably bringing down 
 the average correlation between the cortex and the medium spiny neurons, but it concerns only a small
 fraction of the cortex, and the correlation in the most lateral sections estimated from coronal series only 
 is close to the cortical average.

\subsection{Peaks for the density of cell types are better separated than for correlations}
\begin{figure}
   \subfloat[\label{subfig-2:dummy}]{
      \includegraphics[width=0.9\textwidth]{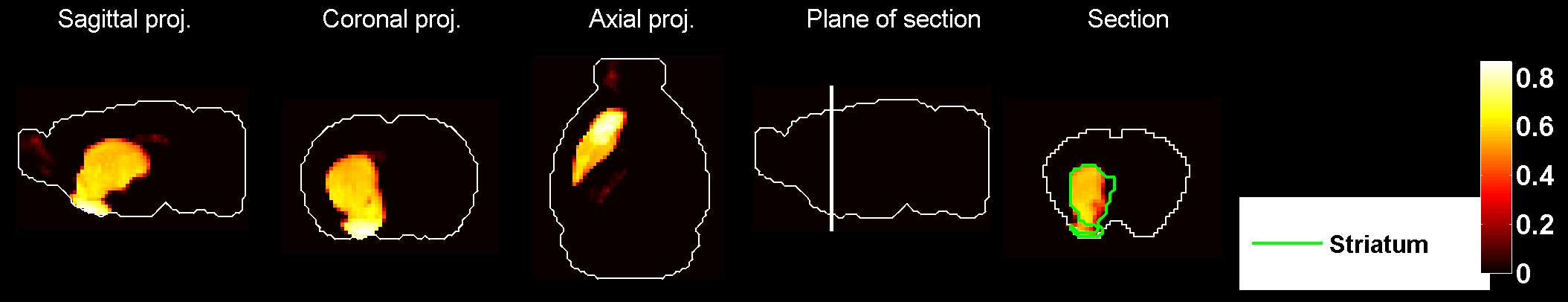}
    }
   \hfill
    \subfloat[\label{subfig-2:dummy}]{
      \includegraphics[width=0.9\textwidth]{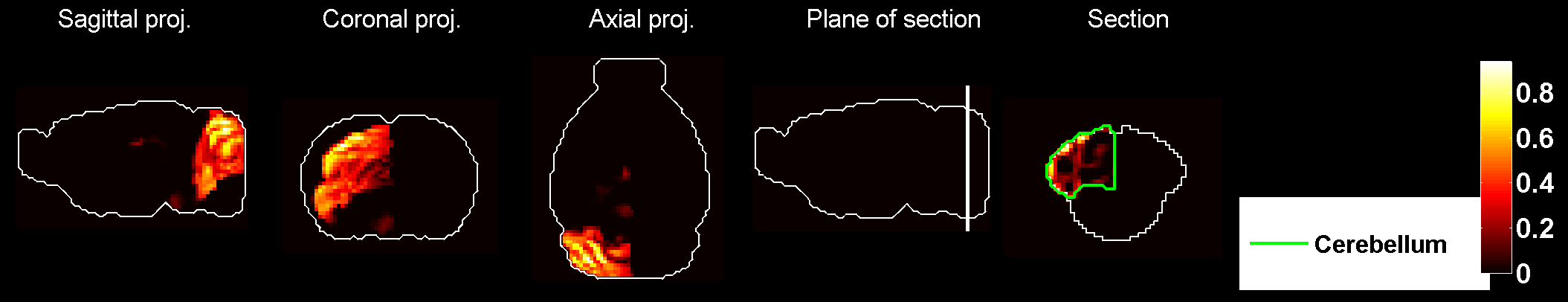}
    }
    \caption{Heat maps of the average density of cell types in the left hemisphere,
 $\langle   \mathrm{Corr} (v,t) \rangle$, defined in Eq. \ref{meanFittingDef}.
 (a) For medium spiny neurons, $t=16$. (b) For granule cells, $t=20$. They are visually very close to
 the heat maps of $\rho_{16}$ and $\rho_{20}$ obtained in \cite{cellTypeBased}.}
    \label{meanFittings}
  \end{figure}

\begin{figure}
   \subfloat[\label{subfig-2:dummy}]{
      \includegraphics[width=0.9\textwidth]{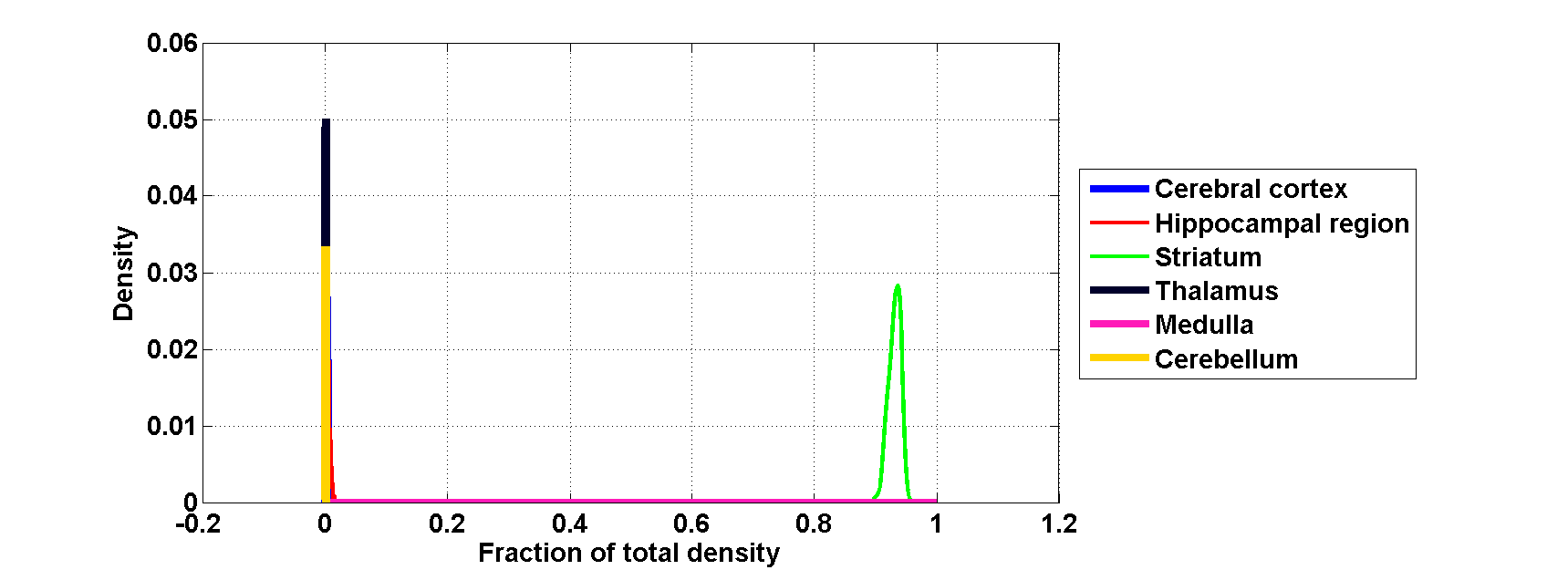}
    }
   \hfill
    \subfloat[\label{subfig-2:dummy}]{
      \includegraphics[width=0.9\textwidth]{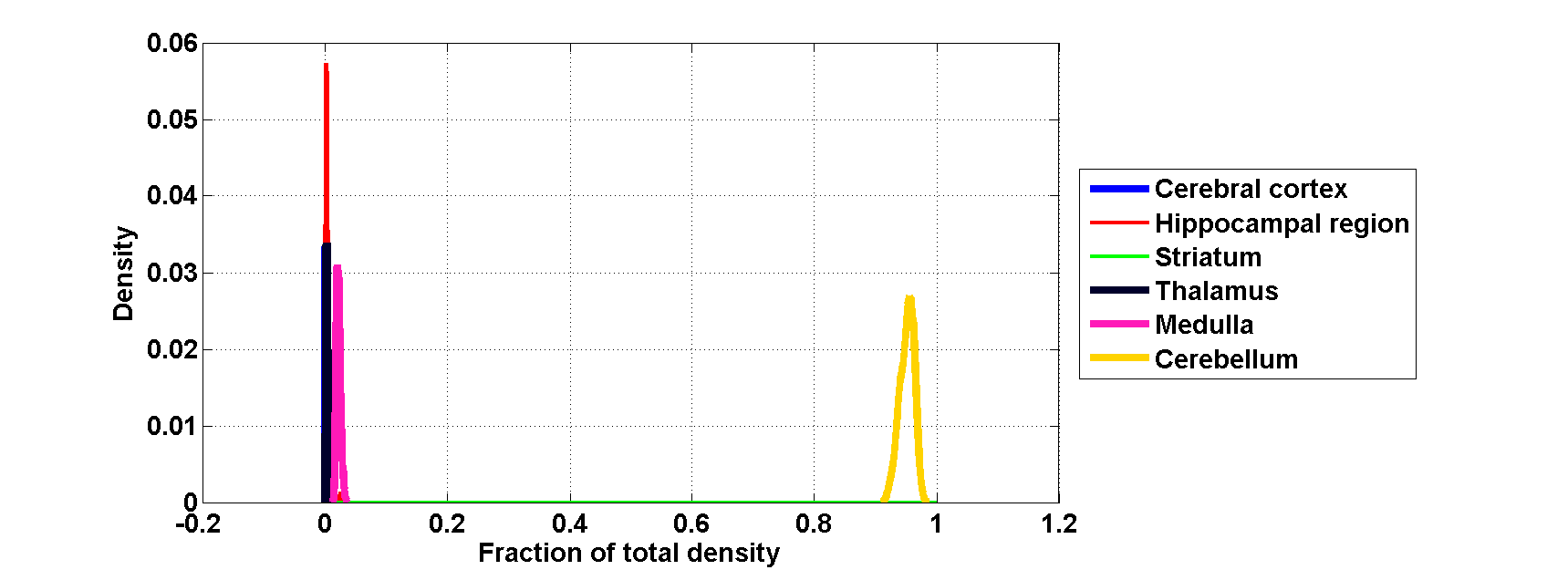}
    }
    \caption{ Estimated probability densities of fractions of density 
 agglomerated in a few regions of the coarsest versions of the ARA (see Eq. 
\ref{fittingDistrDef}). All the region-specific densities
 give rise to peaks. The right-most peak is well-decoupled from the others in both cases, furthermore
 the other peaks are all centered close to zero (making several of them almost invisible).
 (a) For medium spiny neurons, $t=16$; the peak centered at the highest value corresponds to the 
 striatum. (b) For granule cells, $t=20$; the peak centered at the highest value corresponds to the 
 cerebellum.}
    \label{fittingDistrs}
  \end{figure}

Having refitted the linear model (Eq. \ref{linearModel}) for each random draw of image series,
 we can plot the induced average density profiles for the same two cell types as above (see Fig.
 \ref{meanFittings}), which by eye gives a strong impression of the caudoputamen and of the granular layer of the cerebellum in the 
 left hemisphere. Moreover, we estimated the densities of the contribution of each region in the coarsest version 
 of the ARA to the total density of each cell type (Eq. \ref{fittingDistrDef}). The resulting densities are supported in the interval $[0,1]$ 
 by construction, and due to the fact the contributions from each region 
 to the total density sum to 1 in each sample, the righjt-most peaks have a tendency to be more clearly decoupled
 from the others than in the correlation analysis. In particular, we can read from Fig. \ref{fittingDistrs}a that medium spiny neurons
 have $93(\pm3)$ percent supported in the striatum, and from Fig. \ref{fittingDistrs}b similar numbers for granule cells in the cerebellum,
 without any region gathering more than 5 percent of the signal in any of the randoim samples.\\ 


\begin{figure}
 \includegraphics[width=0.9\textwidth]{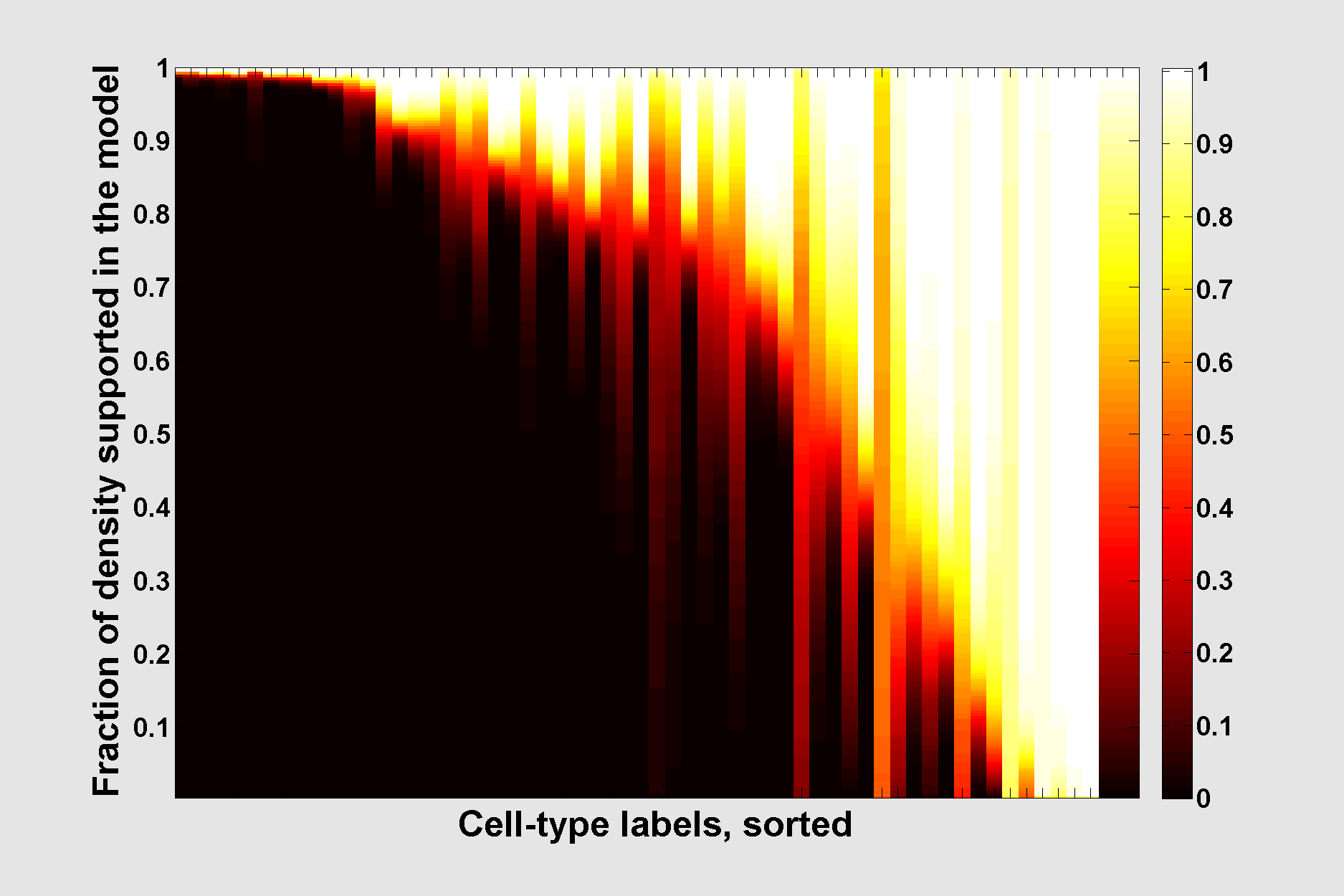}
  \caption{ A heat map of the CDFs of the random overlaps with the results of the model based on the 
 coronal atlas.}
    \label{cdfHeatMapSimulated}
  \end{figure}

 The two cases exposed in Figs. \ref{correlDistrs} and{fittingDistrs} serve 
 as proof of concept, because 
medium spiny neurons and granule cells are well-studied cell types, independently 
 from the neurome approach of \cite{foreBrainTaxonomy}, and they are expected to 
 be strongly associated to the striatum and cerebellum respectively. However, some cell
 types may exhibit more complex neuroanatomical patterns of density, and 
 breaking the estimated density according to the coarsest version of the ARA as in Eq.
 \ref{fittingDistrDef}. To study neuroanatomy in a purely data-driven way, we must not
 use any input from classical neuroanatomy. For a given cell type, we must compare the family of density 
 profiles $(\rho_{[i],t})_{1\leq i \leq R}$ to the density profile $\rho_t$ estimated from the coronal atlas. 
 A possible comparison, proposed in \cite{cellTypeBased} to analyze the results of the 
 sub-sampling simulations, involves the computation of the overlap between density profiles:
\begin{equation}
{\mathcal{I}}(i,t):= \frac{1}{\sum_v \rho_{[i],t}(v)}\sum_v {\mathbf{1}}(\rho_t(v)>0) \rho_{[i],t}(v),
\end{equation}
 which is the fraction of the total estimated density in the $i$-th draw supported by the coronal model.
 For each cell-type label $t$, the overlap ${\mathcal{I}}(.,t)$ is a random variable, whose distribution
 can be studied using the empirical cumulative distribution function (CDF) as follows:
\begin{equation}
 {\mathrm{\sc{CDF}}}( u, t ) = \frac{1}{S}\left|\left\{ i \in [1..R],  {\mathcal{I}}(r,t) \leq u \right\} \right|.
 \end{equation}
 If we present the CDFs in matrix form, with one cell type per column, and rows corresponding to
 the value of the overlap, we can plot this matrix as a heat map, as on Fig. \ref{cdfHeatMapSimulated}. The more
 stable the model is for a cell
 type, the larger the dark area in the corresponding column is. This Figure 
  has a much larger dark area than the analogous Figure in \cite{cellTypeBased}, in which
 the random step involved taking a random 10\% of the genes and refitting the model. For instance, 10 cell types
 have an overlap of more than 95 \% with the original model with probability 1, which was 
 not the case for any of the cell types according to the sub-sampling procedure.
 The computational treatment of the full set of image series therefore 
 reveals stronger stability properties of the linear model.\\

\subsection{Ranking of cell types by stability of results}




Let us recall the notations introduced in \cite{supplementary2} to analyze the results 
 of the simulated distribution of overlaps with the orioginal (coronal) model.\\

Having simulated the distribution of the sub-sampled densities of all the $T$ cell-type
specific transcriptomes in our data set, we can estimate confidence thresholds
 in two ways, for a cell type labeled $t$.\\

 (1) Impose a threshold $\alpha$ in the interval $[0,1]$ 
on the overlap with the density $\rho_t$ estimated in the linear model,
 and work out the probability $p_{t,\alpha}$ of reaching that threshold from the 
sub-samples:\\
\begin{equation}
p_{t,\alpha} := P( \mathcal{I}(.,t) \geq \alpha) = \frac{1}{S}\left| s\in [1..S],   \mathcal{I}(s,t) \geq \alpha\right|.
\label{ptAlpha}
\end{equation}
 For a cell type labeled $t$, the distribution of the overlaps ${\mathcal{I}}(.,t)$ 
 can be visualized using the cumulative distribution function ${\mathrm{\sc{CDF}}}_t$ (in the space $[0,1]$ of the
 values of the overlap between $\rho_t$ sub-sampled profiles $\rho_t^{(s)}$):
 \begin{equation}
 {\mathrm{\sc{CDF}}}_t( \alpha ) = \frac{1}{S}\left|\left\{ s \in [1..S],  {\mathcal{I}}(s,t) \leq \alpha \right\} \right|.
 \end{equation}
 The value ${\mathrm{\sc{CDF}}}_t(u)$ is related 
 to the probability defined in Eq. \ref{ptAlpha} as follows:
\begin{equation}
p_{t,\alpha} = 1 - {\mathrm{\sc{CDF}}}_t( \alpha ).\\
\end{equation}

 (2) Impose a threshold $\beta$ in the interval $[0,1]$ on the fraction of sub-samples, 
and work out which overlap $\mathcal{I}_{thresh}(t,\beta)$ with the estimated density $\rho_t$  is reached by that fraction of the 
sub-samples. The threshold value of the intercept $\mathcal{I}_{thresh}(t,\beta)$ is readily expressed
in terms of the inverse of the cumulative distribution function:
\begin{equation}
\mathcal{I}_{thresh}(t,\beta) = {\mathrm{\sc{CDF}}}_t^{-1}(\beta).
\end{equation} 

 The more stable the prediction $\rho_t$ is again sub-sampling,
 the more concentrated the values of  ${\mathcal{I}}(.,t)$ are at high values (close to 1),
 the slower the take-off of the cumulative function  ${\mathrm{\sc{CDF}}}_t$ is, the lower the
 value of ${\mathrm{\sc{CDF}}}_t( \alpha )$ is,  and the 
 larger the probablity $p_{t,\alpha}$ is (for a fixed value of $\alpha$ in [0,1]).\\
 
For a fixed cell type labeled $t$, the values of $p(t,.)$ and  $\mathcal{I}_{thresh}(t,.)$
can therefore be readily read off from a plot of the cumulative distribution 
 function $\mathrm{\sc{CDF}}$ (this plot is in the $\alpha\beta$ plane in our notations).
 For the sake of visualization of results for all cell types, we constructed 
 the matrix ${\mathcal{P}}$, whose columns correspond to cell types,
 and whose rows correspond to values of the threshold $\alpha$:
\begin{equation}
 {\mathcal{P}}( \alpha, r_t^{signal} ) := {\mathrm{\sc{CDF}}}_t( \alpha ),
\label{cdfHeatMapDef}
\end{equation}
 where the index $r_t^{signal}$ in the l.h.s. means that the 
cell types are ordered by decreasing order of overlap between the
 average sub-sampled profile and the predicted profile.
 If the entries of the matrix $\mathcal{P}$ are plotted 
as a heat map (see Fig. \ref{cdfHeatMapSimulated}), 
 the hot colors will be more concentrated in the left-most 
 part of the image. For a fixed column, the more concentrated 
 the hot colors are in the heat map, the more stable the corresponding cell type is.

\section{Discussion}

The present work allows to bring the power of the online facilities 
 of the ABA to the desktop for a collective analysis of gene-expression 
 energies. The example of the analysis of correlation with microarray data sets
 shows the power of the ABA as a tool of analysis for other data sets. This goes 
 beyond the correlation structure of the ABA itself, and its relation to the 
 classical neuroanatomy of the ARA, that can elready be 
 investigated online using the Anatomic Gene Expression Atlas (AGEA, see \cite{AllenAtlasMol}).\\

  When the coronal ABA was used to analyze other data sets, in particular microarray data
 as in \cite{cellTypeBased}, it was assumed that it reflects an average 
 collective  expression of the genome in the adult mouse brain, showing little sensitivity
 to dynamics and to animal-to-animal variations, even though each expression profile
 comes from a different animal. The fact that correlation profiles allowed
 to guess the anatomical origin of many cell types in \cite{cellTypeBased} provided a self-consistent check 
 of this assumption, but incorporating the different ISH image series into the computational analysis
 allows to control the numerical errors in the correlation profiles.\\

 Moreover, the Monte Carlo approach we took to simulate the 
 variability of region-specificity is much more consistent than the sub-sempling approach 
 taken in \cite{cellTypeBased}. The numerical results show a greater stability
 than in the sub-sampling approach on average.\\

 It should be emphasized that the computations in the present paper only
serve as examples, bringing more accurate answers to the most current questions
 posed by the analysis of the ABA. Ultimately, the distribution of any quantity based on the atlas 
 can be computed exactly and studied at each voxel, without the need to agglomerate  
 voxels into neuroanatomical regions.\\

\clearpage
\section{Tables: cell-type-specific transcriptomes: description, labeling and anatomical origin}
For each of the cell-type-specific samples analyzed in this note, the
following two tables give a brief description of the cell type, the
region from which the samples were extracted according to the
coarsest version of the Allen Reference Atlas, and the
finest region to which it can be assigned according to the data
provided in the studies
\cite{OkatyCells,RossnerCells,CahoyCells,DoyleCells,ChungCells,ArlottaCells,HeimanCells,foreBrainTaxonomy}.
 The indices in the first columns of the tables are the ones refered to as $t$.
\begin{table}
\centering
\begin{tabular}{|m{0.05\textwidth}|m{0.39\textwidth}|m{0.18\textwidth}|m{0.38\textwidth}|}
\hline
\textbf{{\footnotesize{Index}}}&\textbf{Description}&\textbf{{\footnotesize{Region in the ARA (\bigTwelve)}}}&\textbf{Finest label in the ARA}\\\hline
1&Purkinje Cells&Cerebellum&       Cerebellar cortex \\\hline
2&Pyramidal Neurons&Cerebral cortex& Primary motor area; Layer 5 \\\hline
3&Pyramidal Neurons&Cerebral cortex& {\small{Primary somatosensory area; Layer 5}} \\\hline
4&A9 Dopaminergic Neurons&Midbrain& Substantia nigra\_ compact part \\\hline
5&A10 Dopaminergic Neurons&Midbrain& Ventral tegmental area \\\hline
6&Pyramidal Neurons&Cerebral cortex& Cerebral cortex; Layer 5 \\\hline
7&Pyramidal Neurons&Cerebral cortex&Cerebral cortex; Layer 5 \\\hline
8&Pyramidal Neurons&Cerebral cortex&Cerebral cortex; Layer 6\\\hline
9&Mixed Neurons&Cerebral cortex& Cerebral cortex \\\hline
10&{\footnotesize{Motor Neurons, Midbrain Cholinergic Neurons}}&Midbrain& Peduncolopontine nucleus\\\hline
11&Cholinergic Projection Neurons&Pallidum& Pallidum\_ ventral region\\\hline
12&{\footnotesize{Motor Neurons, Cholinergic Interneurons}}&Medulla& Spinal cord\\\hline
13&Cholinergic Neurons&Striatum&Striatum \\\hline
14&Interneurons&Cerebral cortex& Cerebral cortex\\\hline
15&Drd1+ Medium Spiny Neurons&Striatum& Striatum\\\hline
16&Drd2+ Medium Spiny Neurons&Striatum& Striatum\\\hline
17&Golgi Cells&Cerebellum& Cerebellar cortex\\\hline
18&Unipolar Brush cells (some Bergman Glia)&Cerebellum&Cerebellar cortex \\\hline
19&Stellate Basket Cells&Cerebellum& Cerebellar cortex\\\hline
20&Granule Cells&Cerebellum& Cerebellar cortex\\\hline
21&Mature Oligodendrocytes&Cerebellum& Cerebellar cortex\\\hline
22&Mature Oligodendrocytes&Cerebral cortex& Cerebral cortex\\\hline
23&Mixed Oligodendrocytes&Cerebellum& Cerebellar cortex\\\hline
24&Mixed Oligodendrocytes&Cerebral cortex& Cerebral cortex\\\hline
25&Purkinje Cells&Cerebellum& Cerebellar cortex\\\hline
26&Neurons&Cerebral cortex& Cerebral cortex\\\hline
27&Bergman Glia&Cerebellum& Cerebellar cortex\\\hline
28&Astroglia&Cerebellum& Cerebellar cortex\\\hline
29&Astroglia&Cerebral cortex&  Cerebral cortex\\\hline
30&Astrocytes&Cerebral cortex&  Cerebral cortex\\\hline
31&Astrocytes&Cerebral cortex& Cerebral cortex \\\hline
32&Astrocytes&Cerebral cortex& Cerebral cortex \\\hline
33&Mixed Neurons&Cerebral cortex& Cerebral cortex \\\hline
34&Mixed Neurons&Cerebral cortex& Cerebral cortex \\\hline
35&Mature Oligodendrocytes&Cerebral cortex& Cerebral cortex \\\hline
36&Oligodendrocytes&Cerebral cortex&  Cerebral cortex\\\hline
37&Oligodendrocyte Precursors&Cerebral cortex& Cerebral cortex \\\hline
\end{tabular}
\caption{Anatomical origin of the cell-type-specific samples (I).}
\label{metadataAnatomyTable1}
\end{table}

\begin{table}
\centering
\begin{tabular}{|m{0.05\textwidth}|m{0.39\textwidth}|m{0.18\textwidth}|m{0.38\textwidth}|}
\hline
\textbf{{\footnotesize{Index}}}&\textbf{Description}&\textbf{{\footnotesize{Region in the ARA (\bigTwelve)}}}&\textbf{Finest label in the ARA}\\\hline
38&\footnotesize{Pyramidal Neurons, Callosally projecting, P3}&Cerebral cortex& Cerebral cortex \\\hline
39&\footnotesize{Pyramidal Neurons, Callosally projecting, P6}&Cerebral cortex& Cerebral cortex  \\\hline
40&\footnotesize{Pyramidal Neurons, Callosally projecting, P14}&Cerebral cortex& Cerebral cortex\\\hline
41&\footnotesize{Pyramidal Neurons, Corticospinal, P3}&Cerebral cortex& Cerebral cortex \\\hline
42&\footnotesize{Pyramidal Neurons, Corticospinal, P6}&Cerebral cortex& Cerebral cortex \\\hline
43&\footnotesize{Pyramidal Neurons, Corticospinal, P14}&Cerebral cortex& Cerebral cortex \\\hline
44&\footnotesize{Pyramidal Neurons, Corticotectal, P14}&Cerebral cortex& Cerebral cortex \\\hline
45& Pyramidal Neurons&Cerebral cortex& Cerebral cortex, Layer 5 \\\hline
46& Pyramidal Neurons&Cerebral cortex& Cerebral cortex, Layer 5 \\\hline
47& Pyramidal Neurons&Cerebral cortex& {\small{Primary somatosensory area; Layer 5}} \\\hline
48&Pyramidal Neurons&Cerebral cortex& {\small{Prelimbic area and Infralimbic area; Layer 5 (Amygdala)}} \\\hline
49&Pyramidal Neurons&Hippocampal region& Ammon's Horn; Layer 6B\\\hline
50&Pyramidal Neurons&Cerebral cortex& Primary motor area\\\hline
51&{\footnotesize{Tyrosine Hydroxylase Expressing}}&Pons& Pontine central gray\\\hline
52&Purkinje Cells&Cerebellum& Cerebellar cortex \\\hline
53&\footnotesize{Glutamatergic Neuron (not well defined)}&Cerebral cortex& Cerebral cortex; Layer 6B (Amygdala)\\\hline
54&GABAergic Interneurons, VIP+&Cerebral cortex& Prelimbic area and Infralimbic area\\\hline
55&GABAergic Interneurons, VIP+&Cerebral cortex& Primary somatosensory area\\\hline
56&GABAergic Interneurons, SST+&Cerebral cortex&Prelimbic area and Infralimbic area \\\hline
57&GABAergic Interneurons, SST+&Hippocampal region& Ammon's Horn\\\hline
58&GABAergic Interneurons, PV+&Cerebral cortex& Prelimbic area and Infralimbic area\\\hline
59&GABAergic Interneurons, PV+&Thalamus& {\small{Dorsal part of the lateral geniculate complex}}\\\hline
60&GABAergic Interneurons, PV+, P7&Cerebral cortex& Primary somatosensory area \\\hline
61&GABAergic Interneurons, PV+, P10&Cerebral cortex& Primary somatosensory area\\\hline
62&\footnotesize{GABAergic Interneurons, PV+, P13-P15}&Cerebral cortex&Primary somatosensory area \\\hline
63&GABAergic Interneurons, PV+, P25&Cerebral cortex& Primary somatosensory area\\\hline
64&GABAergic Interneurons, PV+&Cerebral cortex& Primary motor area\\\hline
\end{tabular}
\caption{Anatomical origin of the cell-type-specific samples (II).}
\label{metadataAnatomyTable2}
\end{table}

\clearpage
\section{Tables of of rankings of cell types by estimates of overlap with the
 coronal model}
\begin{table}
\begin{tabular}{|l|l|l|l|l|l|}
\hline
\textbf{$r^{signal}_t$}&\textbf{Cell type}&\textbf{Index $t$}&\textbf{$\bar{\mathcal{I}}(t)$, (\%)}&\textbf{$p_{t, 0.75}$, (\%)}&\textbf{$\mathcal{I}_{thresh}(t,0.75)$, (\%)}\\\hline
1&\tiny{Granule Cells}&20&99.5&100&99.2\\\hline
2&\tiny{Purkinje Cells}&1&99.4&100&99.2\\\hline
3&\tiny{Motor Neurons, Cholinergic Interneurons}&12&99.2&100&99.2\\\hline
4&\tiny{Pyramidal Neurons}&47&99.1&100&99.2\\\hline
5&\tiny{Pyramidal Neurons}&49&99.1&100&99.2\\\hline
6&\tiny{Glutamatergic Neuron (not well defined)}&53&99.1&100&99.6\\\hline
7&\tiny{Mature Oligodendrocytes}&35&99&100&98.8\\\hline
8&\tiny{Drd2+ Medium Spiny Neurons}&16&98.9&100&98.8\\\hline
9&\tiny{Tyrosine Hydroxylase Expressing}&51&98.9&100&98.8\\\hline
10&\tiny{Pyramidal Neurons, Callosally projecting, P14}&40&98.3&100&98.4\\\hline
11&\tiny{Pyramidal Neurons}&46&98&100&98.4\\\hline
12&\tiny{Astroglia}&28&97.1&100&98.4\\\hline
13&\tiny{Pyramidal Neurons, Corticotectal, P14}&44&96.9&100&97.7\\\hline
14&\tiny{Pyramidal Neurons}&6&92.4&99.6&95.3\\\hline
15&\tiny{GABAergic Interneurons, PV+}&64&92.2&100&93\\\hline
16&\tiny{GABAergic Interneurons, SST+}&57&91.7&100&93.4\\\hline
17&\tiny{Astrocytes}&31&91.1&99.4&93.4\\\hline
18&\tiny{Purkinje Cells}&52&89.9&93.9&95.3\\\hline
19&\tiny{GABAergic Interneurons, SST+}&56&88.2&95.7&93\\\hline
20&\tiny{GABAergic Interneurons, PV+}&59&87&85.8&94.5\\\hline
21&\tiny{Mature Oligodendrocytes}&21&87&100&88.3\\\hline
22&\tiny{Pyramidal Neurons}&48&86.2&98.9&88.7\\\hline
23&\tiny{A10 Dopaminergic Neurons}&5&85.7&84.1&93.4\\\hline
24&\tiny{Cholinergic Neurons}&13&84.7&96.9&87.5\\\hline
25&\tiny{Pyramidal Neurons}&45&82.8&97.7&85.2\\\hline
26&\tiny{GABAergic Interneurons, VIP+}&55&81.6&75&89.5\\\hline
27&\tiny{Motor Neurons, Midbrain Cholinergic Neurons}&10&80.7&94.1&82.8\\\hline
28&\tiny{A9 Dopaminergic Neurons}&4&80.1&74.8&87.5\\\hline
29&\tiny{Oligodendrocyte Precursors}&37&79.6&67.1&91.8\\\hline
30&\tiny{Mature Oligodendrocytes}&22&79&78.4&81.6\\\hline
31&\tiny{Pyramidal Neurons, Callosally projecting, P3}&38&78.5&70.9&95.7\\\hline
32&\tiny{Astrocytes}&32&78.3&65.5&93\\\hline
\end{tabular}

\caption{Cell types ranked by average overlap with the coronal model (I).}
\label{cellTypeRankingTable1}
\end{table}
\clearpage
\begin{table}
\begin{tabular}{|l|l|l|l|l|l|}
\hline
\textbf{$r^{signal}_t$}&\textbf{Cell type}&\textbf{Index $t$}&\textbf{$\bar{\mathcal{I}}(t)$, (\%)}&\textbf{$p_{t, 0.75}$, (\%)}&\textbf{$\mathcal{I}_{thresh}(t,0.75)$, (\%)}\\\hline
33&\tiny{Stellate Basket Cells}&19&77&64.5&80.1\\\hline
34&\tiny{GABAergic Interneurons, PV+, P7}&60&76.3&57.1&89.5\\\hline
35&\tiny{Mixed Neurons}&34&72.8&50.3&83.2\\\hline
36&\tiny{Pyramidal Neurons, Corticospinal, P14}&43&69.9&48.2&87.1\\\hline
37&\tiny{Pyramidal Neurons}&7&69&18.5&73.4\\\hline
38&\tiny{GABAergic Interneurons, VIP+}&54&67.4&15&72.7\\\hline
39&\tiny{Astrocytes}&30&63.5&9.9&69.1\\\hline
40&\tiny{Pyramidal Neurons, Callosally projecting, P6}&39&56.2&40.5&89.1\\\hline
41&\tiny{Mixed Neurons}&33&56.2&21.5&71.9\\\hline
42&\tiny{Oligodendrocytes}&36&54.3&8.5&63.7\\\hline
43&\tiny{Pyramidal Neurons}&50&44.4&9.7&62.9\\\hline
44&\tiny{Drd1+ Medium Spiny Neurons}&15&42.8&-0.4&47.7\\\hline
45&\tiny{Golgi Cells}&17&40.4&36.8&99.2\\\hline
46&\tiny{Pyramidal Neurons, Corticospinal, P6}&42&32&9.3&48.8\\\hline
47&\tiny{GABAergic Interneurons, PV+, P25}&63&30.4&-0.3&39.1\\\hline
48&\tiny{GABAergic Interneurons, PV+}&58&26.2&0.6&35.5\\\hline
49&\tiny{Unipolar Brush cells (some Bergman Glia)}&18&25&-0.4&31.3\\\hline
50&\tiny{GABAergic Interneurons, PV+, P10}&61&17.9&3.1&29.3\\\hline
51&\tiny{Astroglia}&29&14.1&-0.4&18\\\hline
52&\tiny{Pyramidal Neurons}&8&12.1&0.1&14.1\\\hline
53&\tiny{Pyramidal Neurons}&3&9.6&8&0.4\\\hline
54&\tiny{Neurons}&26&4&0.1&5.9\\\hline
55&\tiny{Mixed Neurons}&9&3.8&3.1&0.4\\\hline
56&\tiny{Bergman Glia}&27&1.1&-0.4&0.4\\\hline
57&\tiny{Pyramidal Neurons}&2&0.3&-0.4&0.4\\\hline
58&\tiny{GABAergic Interneurons, PV+, P13-P15}&62&0&-0.4&0.4\\\hline
59&\tiny{Cholinergic Projection Neurons}&11&0&22.2&71.9\\\hline
60&\tiny{Interneurons}&14&0&22.2&71.9\\\hline
61&\tiny{Mixed Oligodendrocytes}&23&0&22.2&71.9\\\hline
62&\tiny{Mixed Oligodendrocytes}&24&0&22.2&71.9\\\hline
63&\tiny{Purkinje Cells}&25&0&22.2&71.9\\\hline
64&\tiny{Pyramidal Neurons, Corticospinal, P3}&41&0&22.2&71.9\\\hline
\end{tabular}

\caption{Cell types ranked by average overlap with the coronal model (II).}
\label{cellTypeRankingTable2}
\end{table}

\section{Cell-type-specific results}





\multido{\i=1+1}{64}{
 \clearpage
\begin{figure}
\includegraphics[width=1\textwidth,keepaspectratio]{meanCorrels\i.png}
\caption{Heat map of the mean correlation profile for cell type labeled $\i$.}
\label{meanCorrels\i}
\end{figure}
\begin{figure}
\includegraphics[width=0.98\textwidth,keepaspectratio]{correlDistr\i.png}
\caption{Distribution profiles of the correlations between transcriptome profile of  cell type labeled $\i$
 across random sets of ISH expression energy profiles.}
\label{correlDistr\i}
\end{figure}
\begin{figure}
\includegraphics[width=1\textwidth,keepaspectratio]{meanFittings\i.png}
\caption{Heat map of the mean density profile for cell type labeled $\i$.}
\label{meanFittings\i}
\end{figure}
\begin{figure}
\includegraphics[width=0.98\textwidth,keepaspectratio]{fittingDistr\i.png}
\caption{Distribution profiles of the density profile of  cell type labeled $\i$
 across random sets of ISH expression energy profiles.}
\label{fittingDistr\i}
\end{figure}
\clearpage
}


\end{document}